## Aging behavior of Haynes® 282® wrought nickel superalloy subjected to a cold pre-deformation by differential speed rolling

Wojciech Polkowski[1]*, Adelajda Polkowska[1*], Sebastian Lech[2*]

[1]*Łukasiewicz Research Network – Krakow Institute of Technology, Zakopiańska 73 Str., 30-418 Krakow, Poland*

[2] *AGH University of Science and Technology, International Centre of Electron Microscopy for Materials Science and Faculty of Metals Engineering and Industrial Computer Science, Mickiewicza 30 Av., 30-059 Krakow, Poland*

*corresponding author(s) e-mail: wojciech.polkowski@kit.lukasiewicz.gov.pl (W. Polkowski); adelajda.polkowska@kit.lukasiewicz.gov.pl (A. Polkowska); slech@agh.edu.pl (S. Lech)

**Abstract**

The mechanical response of Haynes® 282®, a wrought gamma prime strengthened nickel-based superalloy, is tailored by using both a cold deformation processing and a post-straining heat treatment including a two-step aging. In this work, the effect of the applied cold pre-deformation by conventional equal speed rolling (ESR) or differential speed rolling (DSR) methods on the aging behavior of Haynes® 282® superalloy, is examined for the first time. For this reason, the solid solution annealed alloy was subjected to ~50% nominal cold thickness reduction by either ESR or DSR methods, followed by a two-step heat treatment at 1010°C/2h and 780°C/8h. The material's structural evolution on each processing step was characterized by using scanning electron microscopy/electron backscatter diffraction and transmission electron microscopy methods. Corresponding mechanical properties were examined in static tensile tests performed on non-standard miniaturized specimens. It was found that, the introduction of a shear strain by rolls speed differentiation in the cold-rolling process results in an (I) more efficient strain-hardening of the alloy in the as-deformed state and (II) an intensive grain-refinement upon the post-deformation annealing. Consequently, the DSR provides a higher room temperature mechanical strength obtained in both aging steps, as compared to the non-deformed and conventionally cold-rolled counterparts. Furthermore, noticeable differences in terms of stability of crystallographic orientations for ESRed or DSRed and heat-treated alloy, were noted.

## 1. Introduction

New high strength heat resistant alloys are developed in order to keep up with increasing requirements for devices applied in aviation or energy industry sectors. A good example of such material is Haynes® 282® nickel-based superalloy that was designed to break through the temperature limitations of Inconel 718 and Waspaloy in gas turbine engines [1]. Furthermore, as the ultra-supercritical steam power plants or supercritical $CO_2$ units are now under a fast development, Haynes® 282® is also being considered as serious candidate for the application in the hot sections of such installations [2, 3].

The material is usually provided in the form of sheets or plates in a solid solution annealed (SSA) state in which it exhibits a high susceptibility to a cold forming processing. In order to put the alloy into a thermally stable high strength condition, a two-step post-deformation annealing treatment is applied [1, 4]. The effect of varying heat treatment parameters on the alloy's mechanical properties have been explored in numerous recent works (e.g. [5-9]). On the other hand, a very low attention has been given so far to the impact of a cold-straining processing [10] on structural and mechanical properties evolution upon the further aging treatment of nickel superalloys.

A differential speed rolling (DSR) is a modification of the conventional rolling process in which a differentiation of rotational speed of the upper and lower rolls (controlled by a rolls speed ratio $R$) is applied. Consequently, a through thickness shear strain is imposed to a processed sheet material. Due to a quantitative and qualitative alteration of the strain state, the DSR method is being considered as one of continuous severe plastic deformation (SPD) methods. Its continuous nature provides a number of advantages over hydrostatic SPD techniques in producing high-strength fine-grained materials [11-13]. So far, the DSR has been successfully applied to improve mechanical properties, formability and performance of various metallic sheet materials including e.g. aluminum alloys [14], magnesium alloys [15], iron and low carbon steel [16], intermetallics [17] or even shape memory alloys [18]. At the same time, there are only few works on DSRed nickel or nickel-based superalloys. Previously [19], we have documented that a pure nickel subjected to the high speed ratio DSR (i.e. by using roll speed ratio $R$=4) shows up to 30% higher ultimate tensile strength than its counterpart processed by conventional equal speed rolling. The improved strength was due to an induction of dynamic recrystallization combined with the effect of formation of shear texture components. Im et al.

[20] have also shown that the DSR process is very effective when used to enhance the mechanical properties through the grain refinement of a model Ni-30Cr alloy.

Regarding the effect of cold-rolling prior the aging treatment on precipitation strengthening a number of works have been reported for aluminum [21-23] or copper-based alloys [24, 25], mostly in terms of balancing their mechanical and electrical properties. For example, Li et al. [21] have found that both precipitate strengthening and electrical conductivity of Al–Mg–Si–Sc–Zr alloy increase with increasing the cold-rolling degree before the aging treatment, due to a combined effects of the precipitation hardening and restoration of crystal lattice in recovery/recrystallization processes. Similar findings were given by Naimi et al. [23], who documented that for cold rolled 2024 and 7075 alloys the maximum value of hardness achieved during aging rises when a pre-straining degree is increased. Furthermore, the cold rolling accelerates kinetics of achieving the maximum age-hardening level. Interestingly, Gazizov and Kaibyshev [22], have reported that a pre-deformation may results not only in a quantitative alteration, but also in a change of type and sequence of precipitated phases. They found that in the Al–Cu–Mg–Ag alloy an introduction of strain-induced boundaries can play a role of sites for heterogeneous nucleation of some transient phases that are not observed in the non-deformed material. Analogously, Guo et al. [24] have recently reported that the discontinuous precipitation behavior of Cu–15Ni–8Sn can be significantly accelerated by a prior cold rolling, which can be attributed to the fact that the introduced dislocations could also act as the potential heterogeneous nucleation sites. Consequently, the mechanical and electrical properties are obviously improved while the peak aging time greatly shortens.

In the present work, to our best knowledge the DSR method is used for the first time as the processing step of a commercial wrought nickel-based superalloy. Our particular goal is to examine the effect of applied cold-straining on the structural and mechanical properties evolution in subsequent aging steps of Haynes® 282® nickel superalloy.

## 2. Materials and methods

### 2.1. Material processing

The commercial Haynes® 282® alloy having the nominal chemical composition listed in **Table 1** was provided by Haynes International company (Kokomo, Indiana, USA) in the form of 0.062" (1.6 mm) thickness sheet, in the solid solution annealed (SSA) condition. A good agreement between the nominal and actual chemical compositions was proven by the results of our checks by X-ray fluorescence (XRF) spectroscopy (macroanalysis) and X-ray energy dispersive spectroscopy (EDX) (microanalysis).

**Table 1**. *The chemical composition of Haynes® 282® superalloy [26].*

| Chemical composition, weight % | | | | | | | | | | |
|---|---|---|---|---|---|---|---|---|---|---|
| Ni | Cr | Co | Mo | Ti | Al | Fe | Mn | Si | C | B |
| bal. | 20 | 10 | 8.5 | 2.1 | 1.5 | max. 1.5 | max. 0.3 | max. 0.15 | 0.06 | 0.005 |

The as provided SSA sheets were cut into rectangular bars having a length of 100 mm and a width of 20 mm and then subjected to a cold-rolling forming. The deformation process was carried out by using a hydraulically controlled sexton-type rolling mill equipped with working rolls having equal diameter of 85 mm. The applied cold-deformation process included a 2-pass rolling under dry conditions to a nominal total thickness reduction of 50%. The sample orientation between consecutive rolling passes was change "back and forth", in order to enhance the shear deformation effect [**15**]. The rolling process was performed under a "constant rolling gap" mode, i.e. the rolling position was kept constant upon each rolling pass through an automatically controlled rolling force compensation. Two different cold rolling variants were applied:

- an equal speed rolling (ESR), in which linear velocities of upper ($v_U$=2 m min$^{-1}$) and ($v_B$=2 m min$^{-1}$) and bottom rolls was the same ($R$=1);
- a high ratio DSR, in which linear velocity of the upper ($v_U$=2 m min$^{-1}$) roll was four times higher than that of the bottom one ($v_B$=0.5 m min$^{-1}$) ($R$=4).

After that, the cold-rolled specimens were subjected to a standard heat treatment. Two-step aging at (I) 1010°C/2h + water quenching and (II) 780°C/2h + water quenching, was applied. For the sake of comparison, the non-deformed SSA alloy was also subjected to the aging treatment under the same conditions. The experimental workflow is schematically presented in **Fig. 1**.

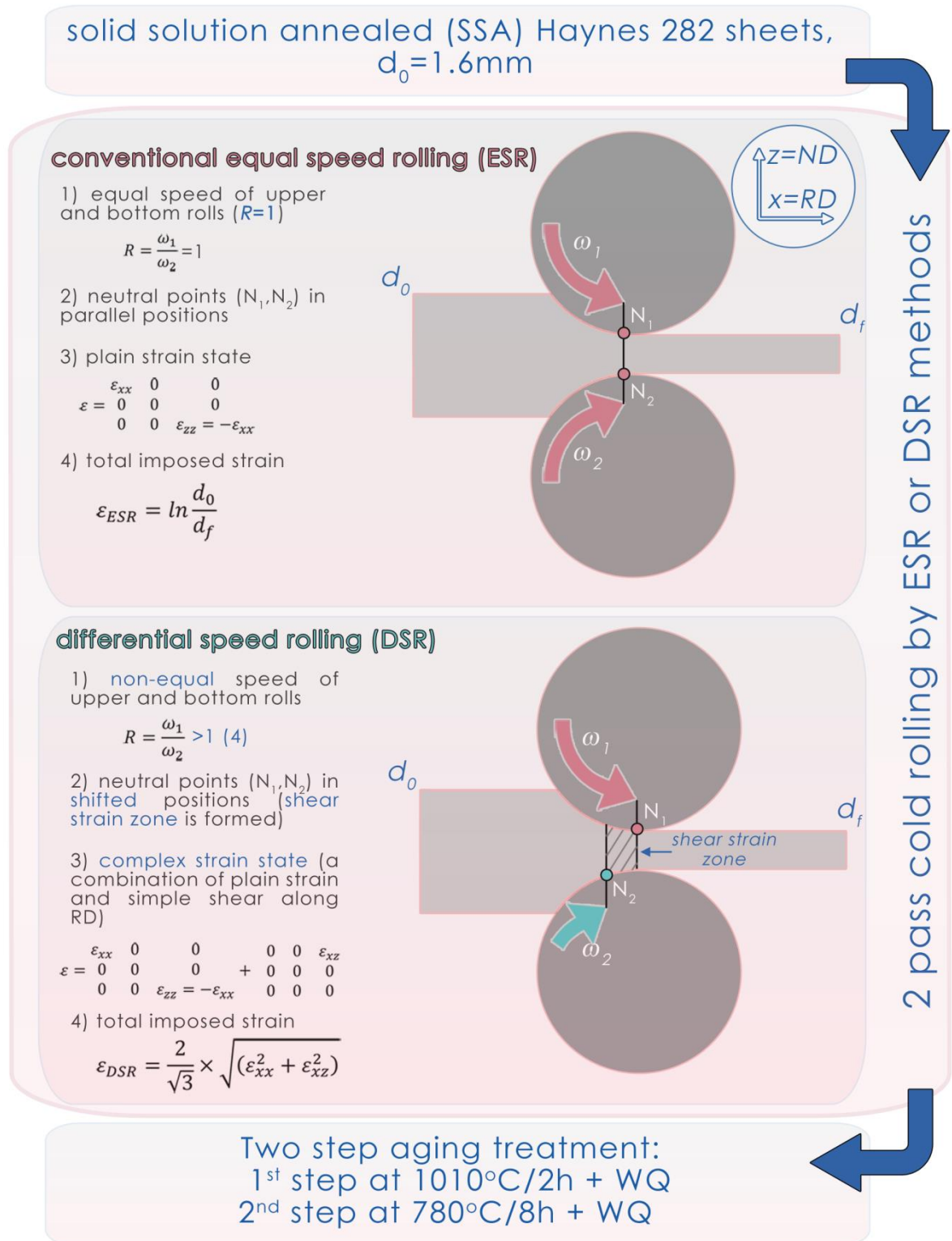

**Fig. 1.** *A scheme showing the experimental workflow including cold-rolling by ESR or DSR and two-stage aging treatment.*

### 2.2. Materials characterization

The microstructure evolution in Haynes® 282® alloy was examined by using light microscopy (LM, Carl Zeiss Axio Observer ZM10) and various electron microscopy techniques. The FEI Scios[TM] field emission gun scanning electron microscope (FEG-SEM) coupled with electron backscatter diffraction (EBSD) system was applied to examine microstructural and microtextural features of processed specimens. The analyses were performed on longitudinal sections of samples, prepared by a mechanical grinding with SiC papers followed by a one hour polishing with silica suspensions. For each sample at least 85000 points were indexed in EBSD measurements. The evolution of crystallographic texture was analyzed by calculating the orientation distribution functions (ODFs) from acquired EBSD data. The harmonic series method was applied upon calculations. The analysis was conducted in an extended Euler angle space ($\varphi_1$=0-360°, $\Phi$=0-90° and $\varphi_2$=0-90°). Transmission electron microscopy (TEM) examinations were performed using Tecnai G2 20 TWIN FEI microscope, operating at the accelerating voltage of 200 kV. Transmission scanning electron microscopy (TSEM) analyses were performed using Merlin Gemini II ZEISS microscope equipped with a Schottky FEG and operating at 30 kV accelerating voltage. For energy-dispersive X-ray spectroscopy (EDXS) a Quantax 800 microanalysis system (Bruker) was used, operated by Esprit 1.9 software. Thin foils were cut off from Haynes® 282® alloy samples and then prepared by using a mechanical polishing followed by a conventional twin-jet electropolishing under 50 V polarization voltage, in a mixture of perchloric acid and glacial acetic acid (1:10) cooled down to 5°C.

Mechanical characteristics of Haynes® 282® alloy were received from room temperature static tensile tests. Non-standard, miniaturized bone-shaped specimens were cut off from cold-rolled sheets along the rolling direction. Static tensile tests at constant strain rate of $6.4 \times 10^{-3}$ s$^{-1}$ were performed by using Zwick Roell AllroundLine Z10 machine equipped with 10 kN load cell and a dedicated specimen grip. The laserXtens HP1-15 high precision (0.5 class) speckle-type laser extensometer combined with optical cameras was applied for non-contact strain measurements. A tensile yield strength (TYS), an ultimate tensile strength (UTS) and a tensile elongation were automatically extracted from recorded curves by a dedicated software (ZwickRoell testXpert II). More details on the used experimental setup are given elsewhere [**10**].

## 3. Results and discussion

### 3.1. Microstructure evolution

A grain structure evolution of Haynes® 282® alloy upon applied thermomechanical processing is shown in SEM images (**Fig. 2**). The material in its as-received SSA condition (**Fig. 2a**) was characterized by an equiaxed grain structure having a mean grain size of 108±14 µm, that was only slightly affected by the applied two-stage aging processing (**Figs. 2b, 2c**). On the other hand, the pre-deformation by cold-rolling produced a strong structural response of the material leading to a formation of banded-like grain structure (**Figs. 2d, g**). What should be noted is the fact that a structure of DSRed material seems to be much more distorted as compared to that of the ESR counterpart, what is also quantitatively reflected as a lower transverse grain boundary spacing (21±16 µm vs. 58±26 µm) as it was determined in SEM/EBSD analyses. This finding is justified by a generally accepted relationship between the imposed true strain and the resulted lamella thickness [**27**]. By taking into account the mathematic model and calculations given by Ko et al. [**28**], the imposed von Misses strain for the ESR (equation 1a) and DSR (equation 1b) processing could be estimated as follows:

$$\varepsilon_{ESR} = \varepsilon_{xx} = ln\frac{d_0}{d_f} \qquad (1a)$$

$$\varepsilon_{DSR} = \frac{2}{\sqrt{3}} \cdot \sqrt{\varepsilon_{xx}^2 + \varepsilon_{xz}^2} \qquad (1b)$$

Where $d_f$ and $d_0$ are reduced and initial thickness, respectively, while $\varepsilon_{xx}$=-$\varepsilon_{zz}$ is a plain strain component and $\varepsilon_{xz}$ is a shear strain component calculated by using the following equations 2a, 2b:

$$\varepsilon_{xz} = \left( \frac{(v_U - v_B) \cdot t_{ave}}{\frac{d_f + d_0}{2}} \right) \cdot \frac{1}{2} \qquad (2a)$$

$$t_{ave} = \sqrt{r(d_0 - d_f) - \left(\frac{d_0 - d_f}{2}\right)^2} \cdot \left(\frac{(v_U + v_B)}{2}\right)^{-1} \qquad (2b)$$

and $v_U$ and $v_B$ are linear velocities of the upper and lower roll respectively, $r$ is the rolls radius and $t_{ave}$ is the average deformation time. By using these assumptions and our experimental conditions, the $\varepsilon_{ESR}$=0.87 and $\varepsilon_{DSR}$=1.39, were obtained respectively.

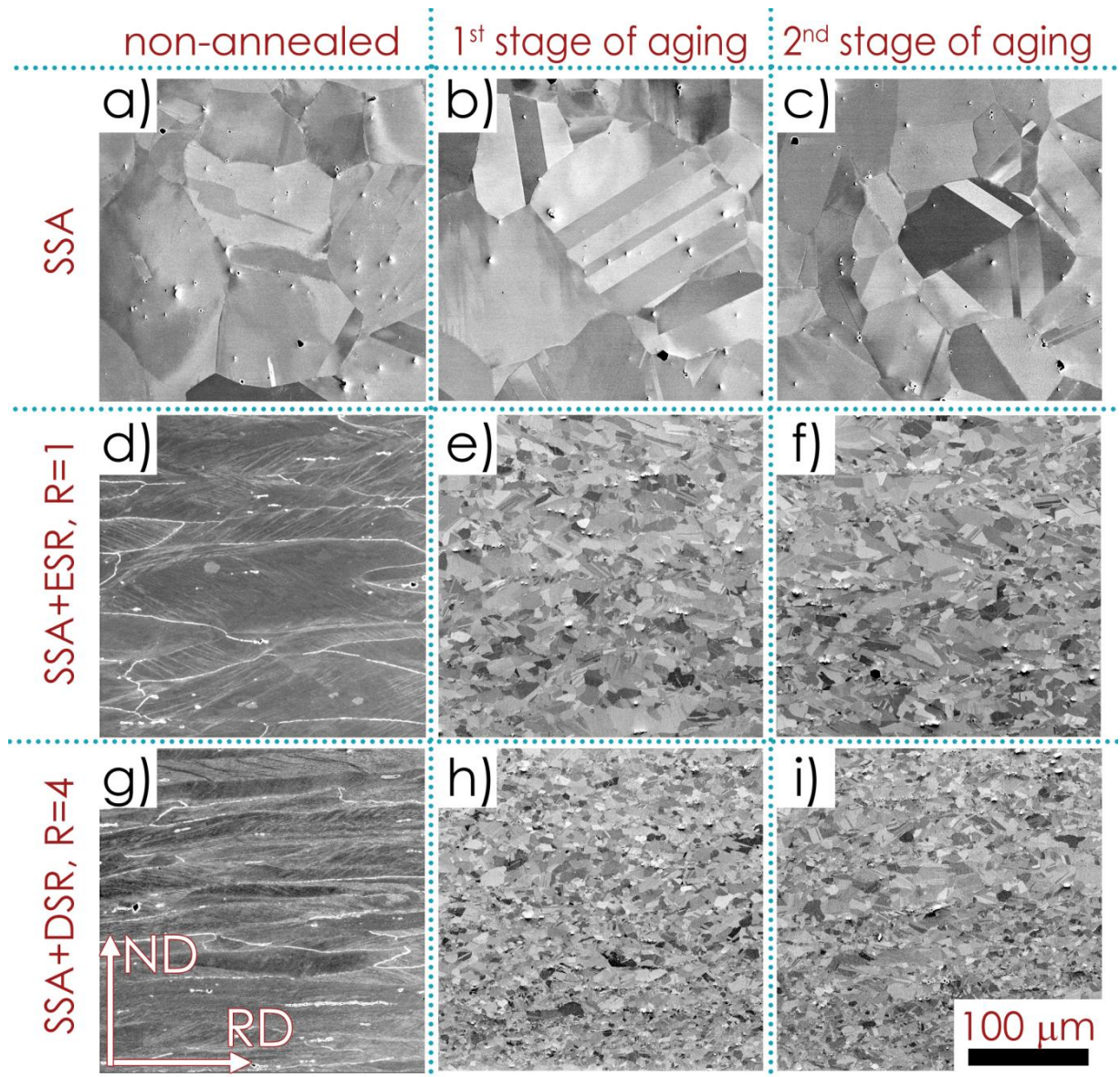

**Fig. 2**. *SEM images showing a grain structure evolution in Haynes® 282® superalloy in SSA state and after two-step aging in SSA conditions (a-c), after the ESR (d-f) and the DSR processing (g-i) (final polishing on silica suspensions, non-etched surfaces).*

As it might have been expected, the pre cold-deformation leads to a substantial grain refinement upon subsequent aging steps. As a consequence of the higher imposed strain, the size of recrystallized grains is visible smaller in DSRed material (**Figs. 2e,f**), as compared to the ESRed counterpart (**Figs. 2h,i**). Interestingly, the light microscopy images taken from chemically etched longitudinal sections (**Fig. 3**) revealed that the banded grain structure of samples cold-rolled by both techniques was maintained after both aging steps. It means that the recrystallization process took place inside deformation bands.

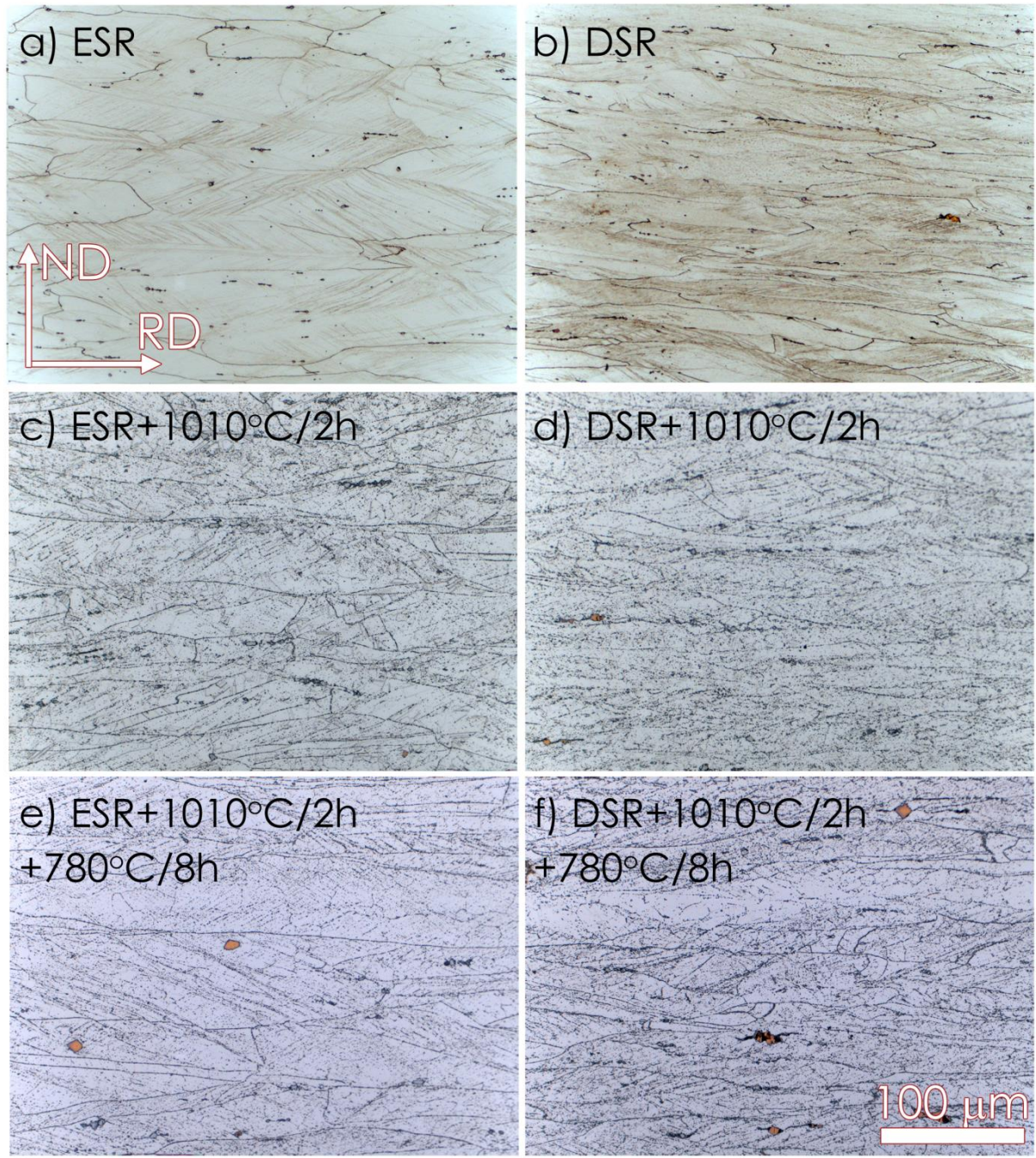

**Fig. 3**. *LM images showing a grain structure evolution in Haynes® 282® superalloy after the cold rolling by either ESR (a) or DSR (b) techniques and after two-step aging (c-f) (a chemical etching with Marble reagent).*

The results of more detailed TEM studies (**Fig. 4**) confirmed a high dislocation density in both cold-deformed and annealed samples. Based on the results of quantitative SEM/EBSD analyses (not shown in here) the size of recrystallized grains for ESD and DSR was found to be 8.6±2.9 μm and 3.5±1.7 μm, respectively. It was reported by Pike [**1**] that during the first aging step at 1010°C/2h fine $M_{23}C_6$ and $M_6C$ carbides are precipitated and stabilized along grain boundaries and in grain volumes, respectively. The main goal of the second step at 780°C/8h is to precipitate γ'-$Ni_3$(Al,Ti) spherical particles having nanoscale dimensions (a mean diameter of 20 nm). Indeed, a massive precipitation of grain boundary carbides (upon the first aging step at 1010°C/2h) and nano-sized $L1_2$ ordered γ' (during the second step at 780°C/8h), was also observed in the present work. Interestingly, it seems that the strain-induced boundaries enhance the carbide precipitation phenomena by providing additional sites for nucleation. Consequently, the banded-like distribution of primary carbides produced upon the cold-deformation was maintained after the full heat-treatment in both types of cold-rolled samples. The smaller lamellar boundary spacing in the DSRed (**Fig. 5a, 5b**) alloy was also kept through a more efficient suppressing of grain boundary motion via the secondary precipitation (in both aging steps). The results of TSEM-EDXS analyses compared with reported literature data [**29, 4, 9**] allows recognizing two types of grain boundary carbides in the fully heat treated material, namely smaller Cr-enriched $M_{23}C_6$ and relatively larger Mo,Ni-enriched $M_6C$ (**Fig. 5c**). The presence of these precipitates is also justified in terms of reported diffusion-based reactions between solid solution and primary MC carbides, assisted by a simultaneous start of the γ' formation. The size of γ' precipitates seems to be not affected by a type of cold-deformation applied before the heat treatment, and in both cases it was around 30 nm (**Fig. 5d**).

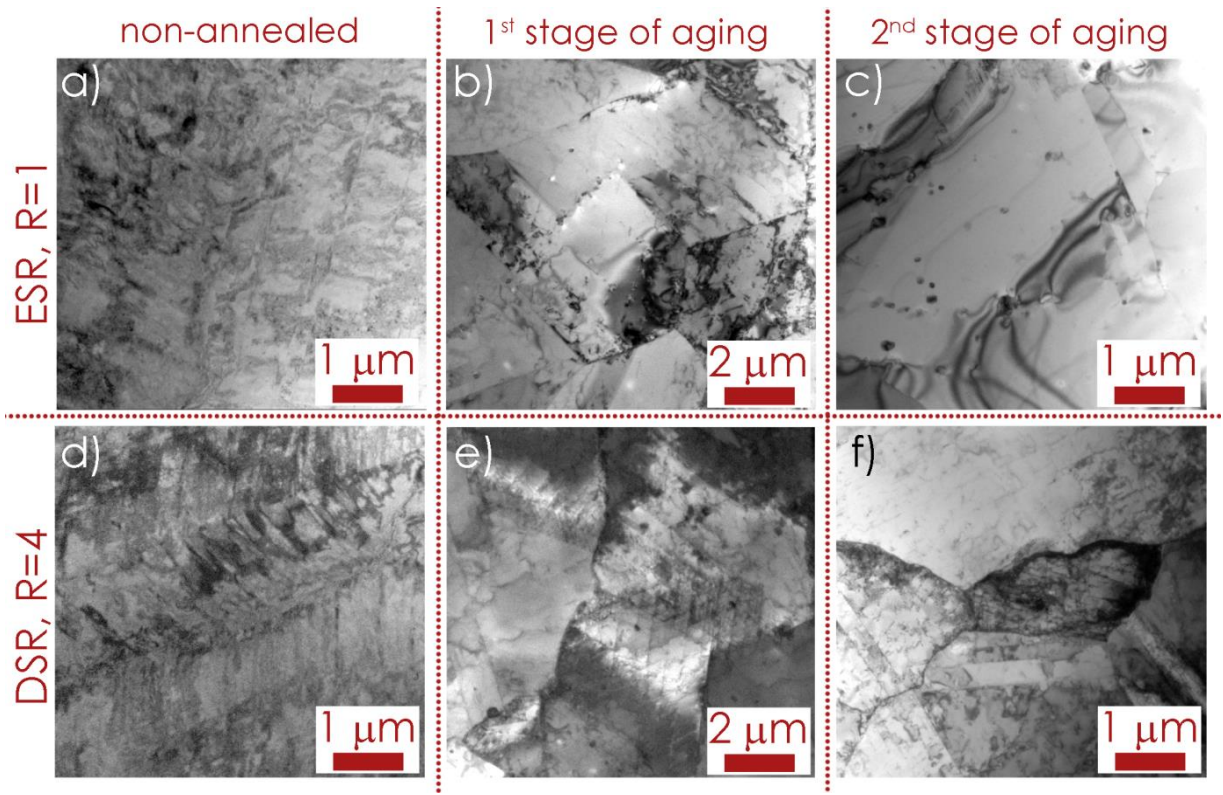

**Fig. 4**. *Bright field TEM images showing a high dislocation density in Haynes® 282® superalloy after the cold rolling by either ESR (a) or DSR (d) techniques and after subsequent two-step aging (b, c) and (e,f), respectively.*

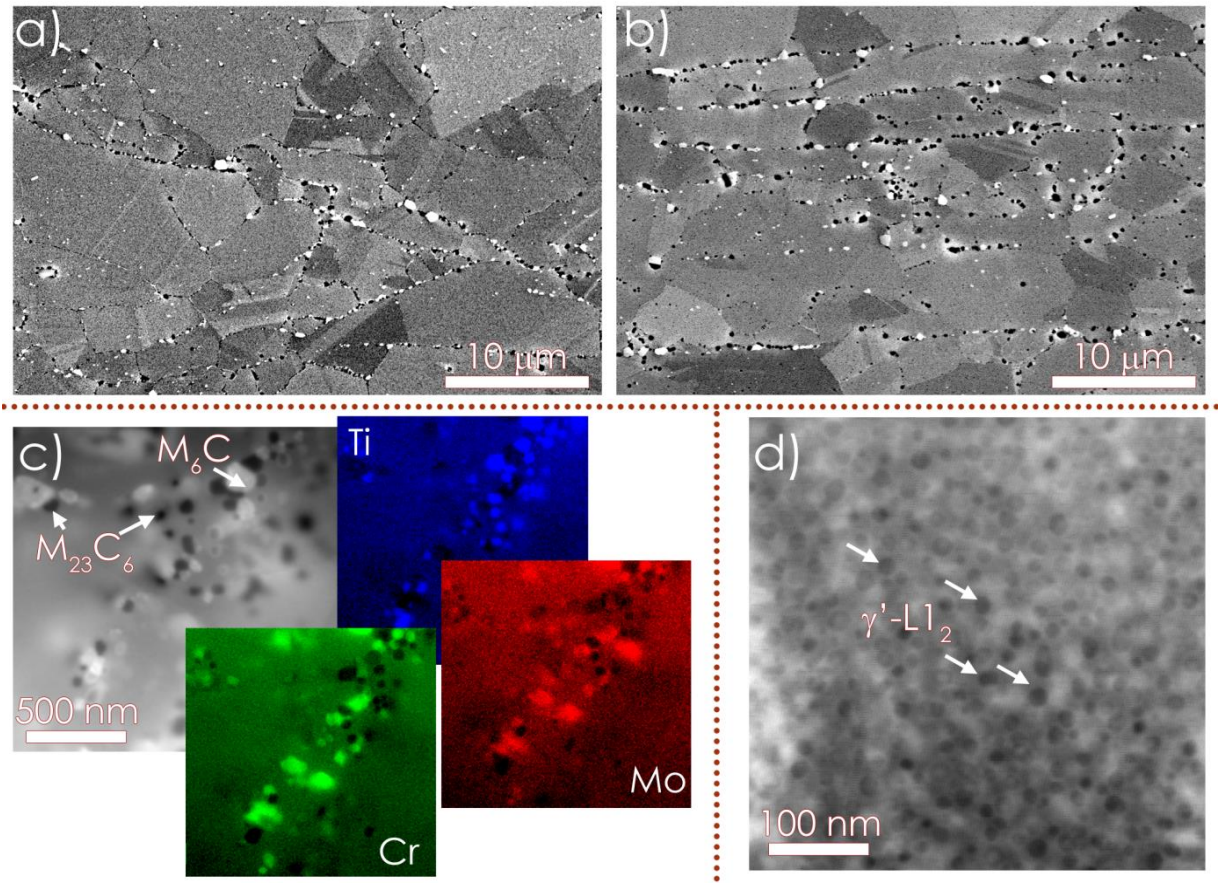

**Fig. 5**. *SEM images of the material after two-step aging subjected a priori to the ESR (a) and DSR (b) processing. The results of TSEM/EDS analyses of grain boundary carbides (c) and nano-sized $L1_2$ ordered γ' particles (d) in the fully heat treated DSRed alloy.*

### 3.2. Crystallographic texture evolution

The ODFs represented in $\varphi_2 = 45°$ sections of the Euler space calculated for the for Haynes® 282® alloy in solid solution annealed, ESRed and DSRed conditions and after 1[st] and 2[nd] stage of the aging treatment, are shown in **Figs. 6a-c**. For the sake of comparison localization of the most important crystallographic orientations in Euler angle space, is given in **Fig. 6d**. The alloy in SSA condition shows a weak recrystallization texture with a slightly pronounced {100}<100> Cube and {100}<110> Rotated Cube components. These features are normally found in fully annealed polycrystalline medium-to-high fcc metals and alloys [**30**]. The ESR processing (**Fig. 6a**) results in a development of deformation texture dominated by {112}<111> Cu and {011}<211> Bs orientations having almost equal intensity. This rolling texture type is typically observed with a high SFE pure fcc metals (e.g. pure nickel), but in the case of highly alloyed (γ+γ'-$L1_2$) nickel superalloys having a relatively lower

SFE [**10**], a contribution of more complex dislocation structure is also taken into account. Moreover, a so called γ-fiber (φ$_1$=0-360°, Φ=54.7° and φ$_2$=45°) connecting shear related {111}<110> and {111}<112> orientations was also produced in the ESRed alloy (**Fig. 6a**) pointing towards a high structure inhomogeneity and strain localization effects.

The implementation of the DSR method results in the strengthening of {112}<111> Cu orientation at the expense of the {011}<211> Bs. Furthermore, a strong "asymmetry" of intensities between Bs$_1$ and Bs$_2$ positions (Fig. 6d) was also observed. It is in line with findings of Clement and Coulomb [**31**], who experimentally documented that the effect of non-equal intensities of Bs orientations in fcc metals is associated with an "asymmetry" of imposed strain.

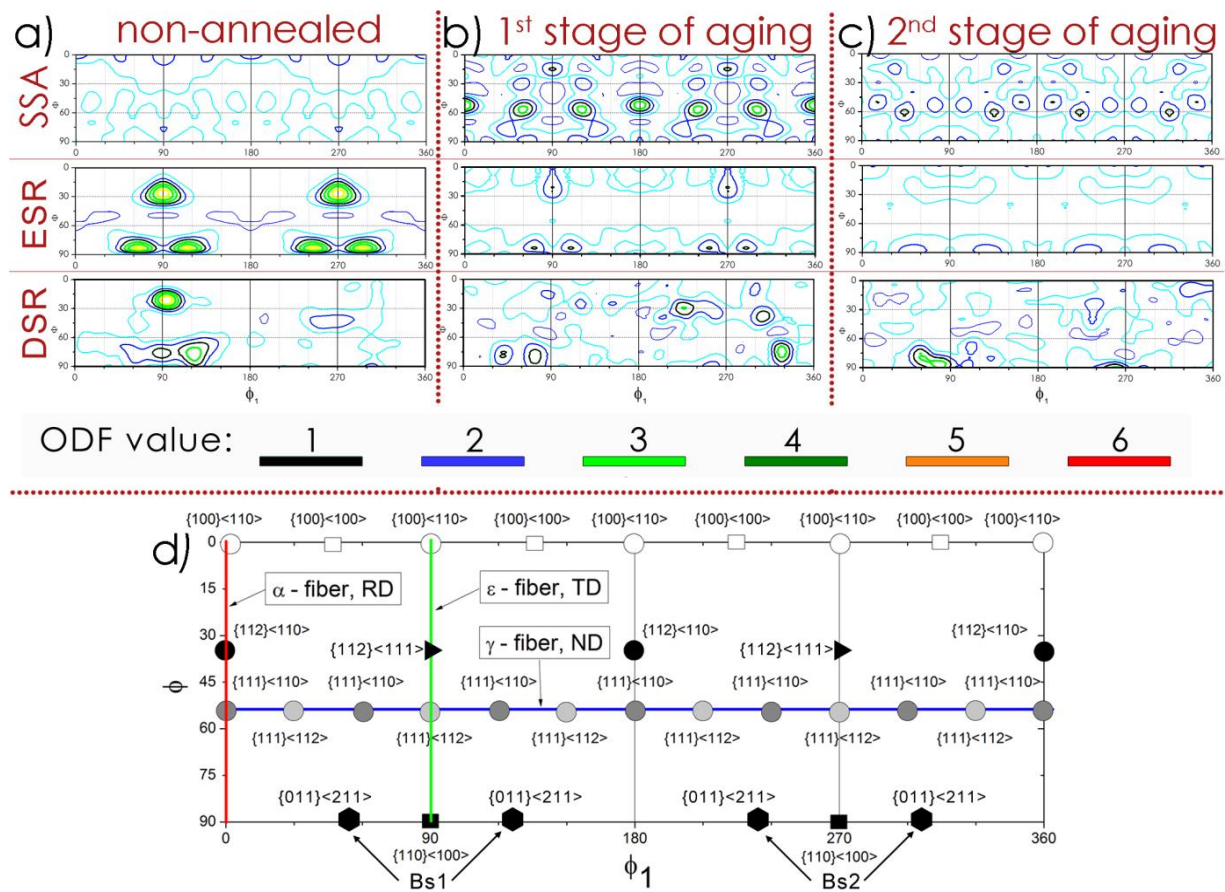

**Fig. 6.** *Crystallographic texture represented in φ$_2$ = 45° sections of the Euler space obtained for Haynes® 282® alloy in solid solution annealed, ESRed and DSRed conditions (a) and after 1$^{st}$ (b) and 2$^{nd}$ (c) stage of the aging treatment. A localization of the most important crystallographic orientations in Euler angle space: recrystallization texture components (white symbols); rolling texture components (black symbols) and shear strain components (grey symbols) (d).*

Nevertheless, from the practical point of view, it is important to show how the applied type of cold deformation processing affects the formability of Haynes® 282® alloy also in a heat-treated condition. An important finding in this matter is that the standard two-step aging of non-deformed SSA alloy (**Fig. 6b**) produced a texture combining the {100}<*uvw*> orientations with a strong {111}<*uvw*> γ-fiber ($\varphi_1$=0-360°, Φ=54.7° and $\varphi_2$=45°) that is very beneficial in terms of formability of fcc alloys [**32, 33**]. As it is found, the conventional ESR processing completely destabilizes these orientations lying along the γ-fiber upon a further heat treatment (they do not appear at all neither after the 1[st] stage nor after the 2[nd] one). As opposite to that, the application of DSR prior the heat treatment gave some measurable traces of the γ–fiber in heat-treated samples.

### 3.3. Mechanical properties

The results of room temperature tensile tests carried out on miniaturized specimens (**Fig. 7a**) taken from Haynes® 282® alloy subjected to various cold-deformation and heat-treatment processes, are shown in **Fig. 7b-d**. The quantitative parameters (TYS, UTS and elongation) are summarized in **Table 2**.

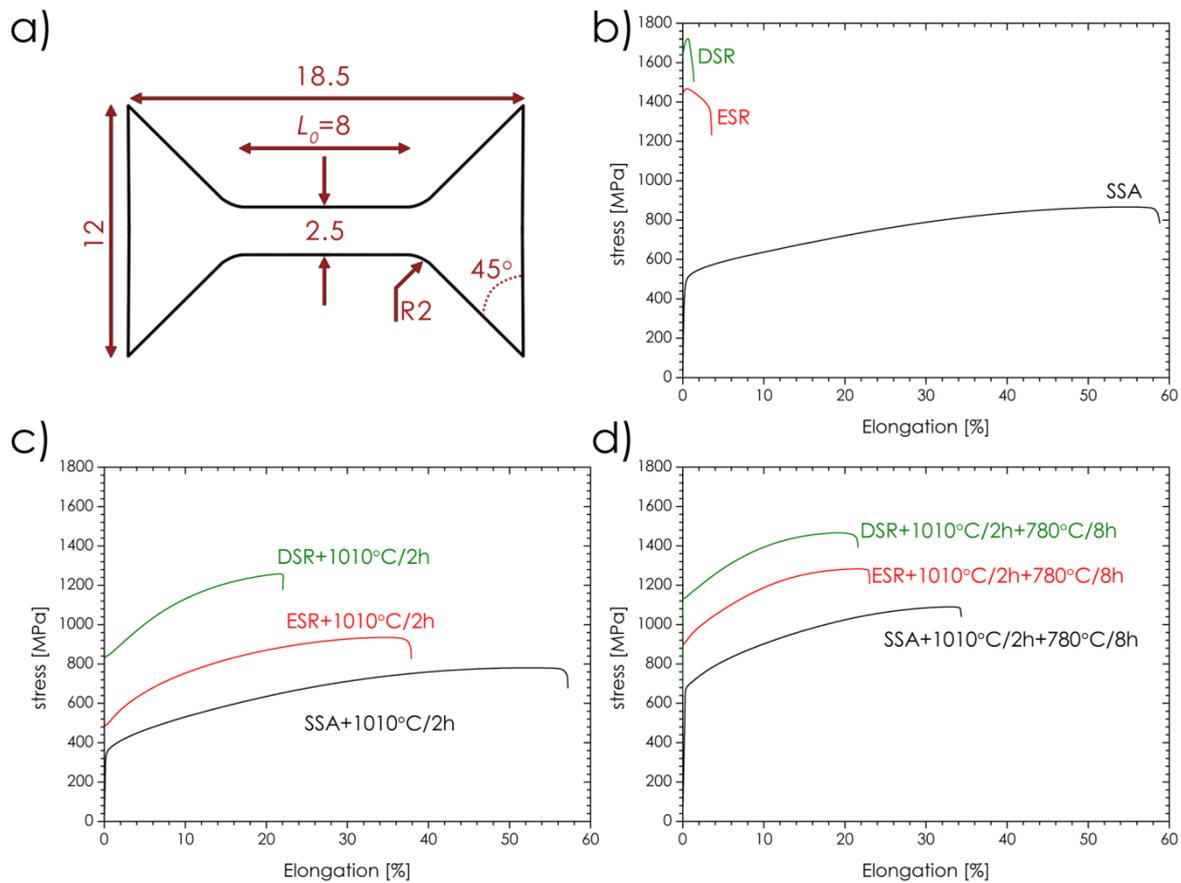

**Fig. 7.** *The results of room temperature static tensile test of Haynes® 282® alloy carried out on miniaturized specimens (a) cut off from SSA and cold deformed materials (b) and then subjected to the first stage aging at 1010°C/2h (c) and the full heat treatment at 1010°C/2h + 780°C/8h (d).*

First of all, it should be noted that the introduction of rolls speed mismatch in the DSR gives much more efficient strain hardening as compared to the ESR (**Fig. 7b**), despite of using the same other cold-rolling parameters. As a consequence of the alteration of imposed strain state (see **Equation 1**) increase of UTS recorded for DSRed alloy is much higher than that of ESRed counterparts ($\Delta UTS_{DSR}$=855 MPa vs. $\Delta UTS_{ESR}$=602 MPa). Analogously, the results of through thickness microhardness measurements points toward high higher and more homogeneous strain hardening in DSRed material (**Fig. 8**). As compared to cold-rolled specimens, an application of the post-deformation annealing results in decreasing alloy's strength (TYS and UTS values) and retrieving a ductility (expressed by increased tensile elongation) of the alloy (**Fig. 7c,d**). Obviously, the course and extent of these changes are strictly driven by a competition between recovery/recrystallization softening and the precipitation strengthening. The higher imposed strain (and the higher stored energy of deformation) in the DSR method results in a faster nucleation rate upon the static recrystallization process, and thus it gives a smaller grain size after both applied post-deformation aging steps. Furthermore, the size of precipitated $L1_2$ - γ' particles seems to be not affected by the type of pre-straining, but apparently higher amount of fine carbides in distributed along GBs and deformation bands in DSRed specimen, was noted. Consequently, after the full heat treatment cycle (**Fig. 7d**), the DSRed alloy showed noticeably higher tensile strength ($UTS_{DSR}$=1489 MPa vs. $UTS_{ESR}$=1282 MPa) than the ESRed counterpart, while having almost the same elongation (21.4 and 23 %, respectively).

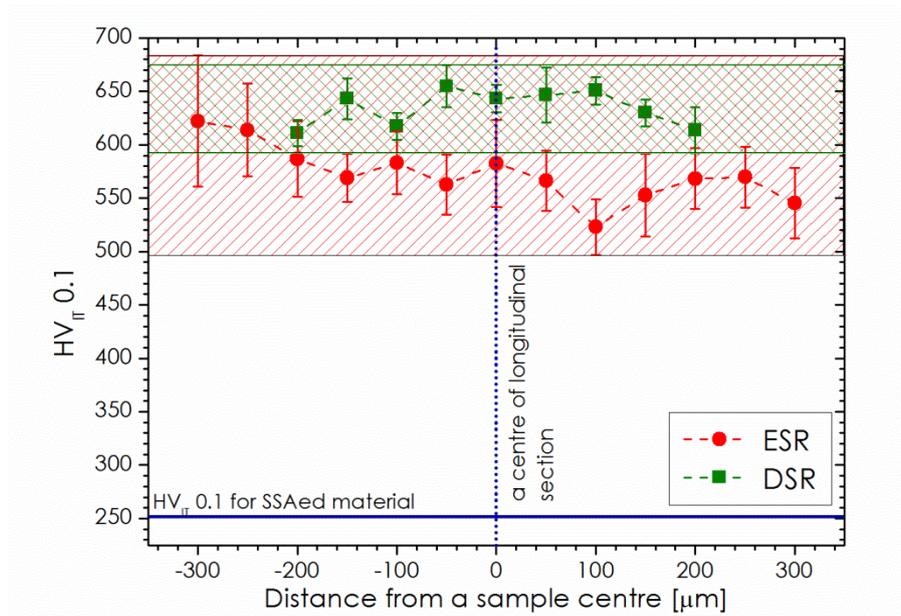

**Fig. 8.** *A through thickness variation of microhardness measured on longitudinal sections of  Haynes® 282® alloy subjected to ESR or DSR cold deformation.*

**Table 2**. The quantitative parameters extracted from the results of room temperature static tensile tests of Haynes® 282® alloy carried out on miniaturized specimens (average values of three specimens).

| Material's condition | Treatment | TYS [MPa] | UTS [MPa] | Elongation [%] |
|---|---|---|---|---|
| as-received | SSA | 398 | 865 | 58.1 |
| | 1st step of aging: 1010°C/2h | 357 | 780 | 57 |
| | 2nd step of aging: 1010°C/2h+780°C/8h | 684 | 1089 | 34 |
| ESR, R=1 | Cold-rolled | 1443 | 1467 | 3.4 |
| | 1st step of aging: 1010°C/2h | 482 | 935 | 37.8 |
| | 2nd step of aging: 10101010°C/2h+780°C/8h | 907 | 1282 | 23 |
| DSR, R=4 | Cold-rolled | 1640 | 1720 | 1.35 |
| | 1st step of aging: 1010°C/2h | 836 | 1257 | 22.1 |
| | 2nd step of aging: 10101010°C/2h+780°C/8h | 1130 | 1489 | 21.4 |

### 3.4. Discussion

In the view of very limited information for nickel-based superalloys, Mei et al. [**34**] have investigated the effect of cold rolling on the precipitation kinetics of intermediate phases in Inconel 718 alloy by using isothermal and non-isothermal calorimetric measurements. They have documented that values of the exothermic effects related to the precipitation of bcc γ'' ($Ni_3Nb$) phase are shifted towards lower temperatures, when the cold rolling degree increases. Furthermore, the pre-straining also promoted a change of the morphology of these particles from disc-shaped towards globular ones. In other words, stored energy of deformation plays a role of an additional driving force for achieving equilibrated structures in shorter time and/or at lower temperatures.

In the view of present work, it is important to rise the following question: *What is the role of the shear-assisted rolling (the DSR processing) in terms of dispersion of particles and reinforcements in metals (and metal matrix composites)?*

Very recently (in 2020), some important answers to aforementioned question were provided by Bahmani and Kim [**35**] in an extensive review study. Based on the reported literature data for various magnesium alloys, e.g. [**36**, **37**], they documented the capability of DSR in refining as-cast eutectic structures and crushing and dispersing the secondary phases, that in turn enhances superplasticity of these materials. Furthermore, some interesting findings have been recently reported by Soeng and Kim [**38**], on the example of Mg-Ca alloys. The authors documented that the additional shear strain imposed in the DSR method followed by a post-rolling annealing leads to a breaking-up clusters of intermetallic phases into fine and isolated particles and their uniform dispersion in the matrix. As a consequence, the material exhibited a more homogeneous grain structure, a better mechanical strength, a superior corrosion resistance and improved thermal stability (a uniformly distributed particles efficiently inhibit undesired grain growth) as compared to counterparts subjected to the processing including the conventional (equal speed) rolling. In the field of metal matrix composites, similar conclusions have been presented for Al/TiC materials by [**39**]. It was shown that a shear strain, which is applied through the DSR, not only refines the grain sizes, but also can effectively disperse the micro- and nanosized reinforcements in the metal matrix composites, which is crucial for the mechanical properties of the material.

Finally, by comparing the results reported in literature with findings of our present work, we may conclude that in the case of Haynes® 282® alloy introducing a rolls-speed differentiation, as compared to a conventional sheet rolling, results in an increased mechanical strength due to:

(I) *a quantitative and qualitative alteration of the strain state.* An increased imposed strain gives a smaller lamellar boundaries spacing in the as-cold rolled condition, that seems to be also maintained after the annealing treatment;

(II) *a smaller size of recrystallized grains and smaller spacing between carbide decorated lamellar boundaries.* The recrystallization process seems to take place at a temperature range of the first stage aging treatment, so most probably simultaneously or just after a "stabilization" and precipitation of GB carbides. The nucleation and growth of new grains occurs within deformation bands, while $M_6C$ and $M_{23}C_6$ carbides precipitated along lamellar boundaries strongly reduces the high angle grain boundary mobility needed for the recrystallized grains growth. Thus, a higher tensile strength of DSRed and heat treated alloy might be justified by both a smaller grain size (according to the well-known Hall-Patch relationship) and a smaller spacing between lamellar boundaries decorated with precipitated carbides.

### 4. Conclusions and future remarks

In the present work, the effect of high ratio differential speed rolling on structure/mechanical response and aging behavior of Haynes® 282® alloy, was experimentally examined for the first time. Based on the obtained results and reported literature data, it is found that, as compared to conventional equal speed rolling, an additional shear strain imposed in the DSR processing enhances the room temperature mechanical strength of the alloy in it as-cold deformed condition, as well as upon the full standard heat treatment (two step aging). In the former case, it might be associated with the strain hardening due to a quantitatively higher accumulated strain ($\varepsilon_{DSR}$=1.39 vs. $\varepsilon_{ESR}$=0.87), what is also reflected by a lower lamellar boundary spacing. In the latter case, the higher stored energy of deformation in the DSR method provides a faster nucleation rate upon the recovery and recrystallization processes, and thus – a smaller grain size after the full heat treatment. Furthermore, a massive precipitation of $M_6C$ and $M_{23}C_6$ carbides along strain-produced lamellar boundaries allows maintaining the "banded" structure and distribution of precipitates even after both aging steps. In the case of ESRed specimens, it was reflected by a total destabilization of the {111}<*uvw*> γ-fiber, that is desirable from the point of materials deep drawing susceptibility, On the other hand, these orientations were still observed in DSRed and annealed samples.

Finally, we believe that a modification of rolls speed mismatch upon the processing of Haynes® 282® alloy might be used to give a further improvement of performance properties of this very good material. Nevertheless, some future works are needed, especially, in terms of examining the effect of altering a rolls speed ratio and the aging heat treatment parameters, as well as to determine the behavior of DSRed Haynes® 282® sheets under predicted working conditions.

### Acknowledgements

A financial support from the Polish National Science Centre under Grant no. UMO-2016/23/D/ST8/01269 (SONATA 12) is gratefully acknowledged. The work has also received a partial financial support from the Łukasiewicz Research Network Foundry Research Institute within its own activities (project no: 9603/00) founded by Polish Ministry of Science and Higher Education in 2019. A part of the study was performed within AGH statutory project No. 16.16.110.663 and supported by the infrastructure of the International Centre of Electron Microscopy for Materials Science (IC-EM), AGH-UST The authors would like to express their acknowledgments to Mr. Robert M. Purgert (Energy Industries of Ohio, USA) and Dr. Paweł Jóźwik (Military University of Technology, Poland) for their assistance in handling the materials and performing the plastic deformation processing, respectively.

**Data availability statement**

The raw data required to reproduce these findings are stored on laboratory computers and are available on-demand from Authors.